\documentclass[aps,prc,twocolumn]{revtex4}
\usepackage{amsmath,amssymb}
\usepackage{graphicx}
\usepackage{color}
\usepackage{mathrsfs}

\begin{document}

\title{\boldmath SS-HORSE Extension of the No-Core Shell Model: Application
to Resonances in  $^7$He }
\author{I. A. Mazur}
\affiliation{Center for Extreme Nuclear Matters, Korea University, Seoul 02841, Republic of Korea}
\affiliation{Laboratory for Modeling of Quantum Processes, Pacific National University, Khabarovsk 680035, Russia}
\author{I. J. Shin}
\affiliation{Rare Isotope Science Project, Institute for Basic Science, Daejeon 
    { 34000}, Republic of Korea}
\author{Y. Kim}
\affiliation{Rare Isotope Science Project, Institute for Basic Science, Daejeon 
    { 34000}, Republic of Korea}
\author{A. I. Mazur}
\affiliation{Laboratory for Modeling of Quantum Processes, Pacific National University, Khabarovsk 680035, Russia}
\author{A. M. Shirokov}
\affiliation{Skobeltsyn Institute of Nuclear Physics, Lomonosov Moscow State University, Moscow 119991, Russia}
\author{P. Maris}
\affiliation{Department of Physics and Astronomy, Iowa State University, Ames, IA 50011-3160, USA}
\author{J. P. Vary}
\affiliation{Department of Physics and Astronomy, Iowa State University, Ames, IA 50011-3160, USA}

\begin{abstract}

Theoretical {\it ab initio} studies of resonances in the unbound  $^7$He nucleus are presented. We
perform no-core shell model calculations with $NN$ interactions Daejeon16 and JISP16 and utilize 
the \mbox{SS-HORSE} method  to calculate the $S$ matrix for two-body channels $n{-}{\rm^{6}He}$ and
$n{-}{\rm^{6}He^{*}}$ with $^{6}$He respectively in the ground and excited $2^{+}$ states as well as
for the four-body democratic decay channel ${{\rm^{4}He}+n+n+n}$. The resonant energies and widths are
obtained by numerical location of the $S$-matrix poles. We describe all experimentally known $^{7}$He
resonances and suggest an interpretation of an observed wide resonance of unknown spin-parity.
\end{abstract}


\maketitle

\section{Introduction}

A modern trend of nuclear theory is the development of methods for 
describing nuclear states in the continuum, resonances in 
particular, and the boundaries of nuclear stability 
as either neutron number or proton number is increased to 
the point where the nucleus becomes unbound.
The $^7$He nucleus presents an 
especially significant challenge since it has no bound states and 
the experimental information on its resonances is fragmentary. 
Ideally, an approach with predictive power could help refine 
current knowledge of $^7$He and
inform further experimental efforts. For 
maximal predictive power, {\it ab initio} (``first-principles'') 
approaches in this field are of primary importance since the
only input is the interaction between nucleons.

Currently there are a number of reliable methods for the
{\it ab initio} description of nuclear bound states (see, e.\:g., 
the review~\cite{Leidemann}). Prominent methods include
the Green function's Monte Carlo~\cite{GFMC}, the no-core shell 
model (NCSM)~\cite{NCSM}, the coupled cluster method~\cite{CCM}, 
etc. The NCSM employed here
is a modern version of the nuclear shell model which does not 
introduce an inert core and includes the degrees of freedom of 
all nucleons of a given nucleus. The multi-particle wave function 
is expanded in a series of basis many-body oscillator functions 
(Slater determinants) which include all many-body oscillator 
states with total number of excitation quanta above the minimum needed to satisfy the Pauli principle that are less than or equal to some given 
value $N_{\max}$. This makes it possible to separate 
the center-of-mass motion. 
The degree of convergence achieved with
NCSM calculations 
as~$N_{\max}$ and/or number of nucleons~$A$ increases is
governed by the limits of available
supercomputers.

The NCSM cannot be directly applied to the description 
of resonant states. Energies of resonant states are positive with 
respect to some breakup threshold so that one needs to 
consider decay modes. Special methods taking into account the 
continuum are therefore needed for the description of resonances.

There are well-developed methods for {\it ab initio} descriptions 
of continuum spectrum states based on Faddeev and 
Faddeev--Yakubovsky equations that are successfully applied to
systems with $A\leq5$ nucleons (see, e.\:g., 
the review~\cite{Leidemann} and Ref.~\cite{Rimantas}). A very 
important breakthrough in developing {\it ab initio} theory 
of low-energy
reactions in heavier systems 
was achieved by combining the NCSM and the resonating 
group method to build the NCSM with continuum (NCSMC) 
approach~\cite{Navratil} which has been applied to nuclear 
systems up to $A=12$~\cite{Navratil_Be11, talkPetr}. Nuclear 
resonances can also be obtained in the no-core Gamow shell model 
(GSM)~\cite{NCGSM}. However, these methods provide significant 
numerical challenges for no-core 
systems~\mbox{\cite{NCGSM,Shin:2016poa,JohnsonJPG2020}}. At higher energies, 
above the resonance region, alternative {\it ab initio} methods 
are developed and applied (see, e.g., Ref. 
\cite{Burrows:2018ggt}).

Recently we proposed the SS-HORSE method~\cite{SSHORSE, 
SSHORSE_PEPAN,Blkh,Blokh, SSHORSE-K,PEPAN2}, which 
generalizes the NCSM to the continuum 
states. The 
SS-HORSE allows one to calculate the single-channel $S$-matrix 
and resonances by a simple analysis of NCSM eigenenergy behavior 
as a function of parameters of the many-body oscillator basis. 
The SS-HORSE extension of the NCSM was successfully applied
to the calculation of the neutron--$\alpha$ and proton--$\alpha$  
scattering and resonant states in the $^{5}$He and $^{5}$Li 
nuclei in Refs.~\cite{SSHORSE, SSHORSE-K}; a generalization of 
this approach to the case of the democratic decay provided a 
prediction of a resonance in the system of four neutrons 
(tetraneutron)~\cite{tetran, Shirokov:2018und} 
whose first low-statistics observation~\cite{tetran-exp} has been followed by its discovery and characterization in a high-statistics experiment~\cite{nature}.

The unbound $^{7}$He nucleus presents a new challenge for 
{\it ab initio} theory but is especially interesting since its
experimental information is fragmentary and
conflicting. A few resonances have been observed in $^7$He 
but all
have { weak} spin-parity assignments  if 
any~\cite{Tilley}. In particular, the lowest resonance
with a
width of~0.18~MeV 
at the energy of 
0.43~MeV above the ${n+{\rm^{6}He}}$ threshold~\cite{Cao} has a tentative spin-parity of $3/2^{-}$. There 
is also a resonance at~3.36~MeV with the width of~1.99~MeV 
which is tentatively assigned~${J^{\pi}=5/2^{-}}$ and another
resonance at~$6.2\pm 0.3$~MeV with the width of~$4\pm1$~MeV of 
unknown spin-parity~\cite{Tilley}. The most complicated situation 
is with the~$1/2^{-}$ resonance which was observed in 
Refs.~\cite{Wuosmaa, Boutachkov, Meister}: according to these works,
its energy ranges from~1~\cite {Meister}
to~3.5~MeV~\cite{Boutachkov} and the width 
from~0.75~\cite {Meister} to~10~MeV~\cite{Boutachkov}.
Thus, $^{7}$He represents a very good candidate for 
invoking
the predictive power of {\em ab initio} scattering theory.
Therefore, we predict the 
 resonances of $^7$He 
within the SS-HORSE--NCSM approach. We find additional broad 
resonances that 
{suggest a new interpretation of $^{7}$He resonant structure.}

Recent many-body calculations of the $^{7}$He nucleus, explicitly 
accounting for the continuum spectrum effects, include a GSM study 
of Ref.~\cite{GSM},  Gamow-density-matrix renormalization-group (G-DMRG)
calculations of Ref.~\cite{DMRG}, an investigation within the 
complex-scaled cluster-orbital shell model (CS-COSM) in Ref.~\cite{CS-COSM},  a NCSMC study of
Refs.~\cite{Navratil7He1,Navratil7He2}, and  recent calculations of Ref.~\cite{Rodkin-Tchu} which we shall
refer to as NCSMch where the NCSM
wave functions of $^{7}$He are matched with the   wave functions in a particular decay channel.
The \mbox{G-DMRG} approach is based on 
the GSM
but utilizes the many-body technique of the density matrix renormalization 
group~\cite{DMRG1,DMRG2} to speed up the convergence.
The GSM~\cite{GSM},  G-DMRG~\cite{DMRG} 
and CS-COSM~\cite{CS-COSM}
calculations  are performed with the $^{4}$He core and  
nucleons in the $psdf$  (GSM) or $spd$ (G-DMRG and CS-COSM) \mbox{valence} spaces interacting by 
phenomenological effective potentials.
 The \mbox{NCSMC} calculations
of Refs.~\cite{Navratil7He1, Navratil7He2} use a Similarity 
Renormalization Group (SRG)-evolved
chiral next-to-next-to-next-to-leading order (N3LO) 
nucleon-nucleon ($NN$) potential 
of Refs.~\cite{EM1,EM2} while the NCSM calculations of Ref.~\cite{Rodkin-Tchu} utilize the
Daejeon16 $NN$ interaction~\cite{Daejeon16} originating from the same chiral N3LO interaction and 
adjusted with unitary transformations that preserve the $NN$ phase shifts
 to describe accurately binding energies and
spectra of $p$-shell nuclei without three-nucleon ($NNN$) forces.

We present here  {\em ab initio} SS-HORSE--NCSM calculations  
performed using the code MFDn \cite{cpe.3129,SHAO20181}
with  realistic Daejeon16~\cite{Daejeon16}  and JISP16~\cite{JISP16}
$NN$ interactions. The difference with the approach presented in Ref.~\cite{Rodkin-Tchu} 
is that we obtain the resonant energies and widths by locating the $S$-matrix poles. 
The $S$-matrix elements in all channels have the poles at the same location in the
complex energy plane. 
Therefore our resonance widths are the total widths of  resonances associated with decay in all possible channels; 
they may be very different from the partial widths which are obtained within NCSMch~\cite{Rodkin-Tchu} 
or within the NCSMC~\cite{Navratil7He1, Navratil7He2} that
characterize the probability of the decay in a particular
channel.

Our  SS-HORSE--NCSM approach is sketched  in Section~\ref{Sec:sshorse}. Results of
our calculations of  $^{7}$He resonances are presented in Section~III. Section~IV includes summary and
conclusions.

\section{SS-HORSE--NCSM approach\label{Sec:sshorse}}

Our {approach to obtaining} 
resonance parameters is to locate $S$-matrix poles. 
The $S$-matrix in the channel with orbital momentum~$\ell$, 
${S_{\ell}=e^{2i\delta_{\ell}}}$, can be expressed 
 through the effective range function,
\begin{equation}
\label{EffRadius}
K_\ell(E)=k^{2\ell+1}\cot\delta_\ell(E),
\end{equation}
where~$\delta_\ell(E)$ is the phase shift, 
$E$ is the energy of relative motion in a given channel and~$k=\sqrt{2\mu E}/\hbar$
is the relative momentum while~$\mu$ is the reduced mass.
The effective range function~\eqref{EffRadius} has good analytical properties and may be expanded in a power 
series of the energy~$E$ (the so-called effective range expansion).  The  function~$K_\ell(E)$
within the SS-HORSE method is calculated at the eigenenergies of relative motion~$E_{i}$
of a decaying resonant state
obtained in the NCSM  as~\mbox{\cite{SSHORSE, SSHORSE_PEPAN, Blkh}} 
\begin{equation}
\label{PhaseSSHORSE}
K_\ell(E_{i})=
-k_{i}^{2\ell+1}\,\frac{C_{{\mathbb N}^{i}+2, \ell}(E_{i})}
      {S_{{\mathbb N}^{i}+2, \ell}(E_{i})}.
\end{equation}
Here $S_{n,\ell}(E)$ and $C_{n,\ell}(E)$ are the 
regular and irregular solutions of the free Hamiltonian in the oscillator representation
for which analytical expressions in the case of two-body channels can be found in Refs.~\cite{yamani,Bang} 
and in the case of democratic four-body decay channels in Refs.~\cite{zayMPQT,zaytmf}; 
${\mathbb N}^{i}$ is the maximal oscillator quanta of the relative motion in the decaying channel
allowed in the NCSM calculation for $^{7}$He.
Note, the functions~$S_{{\mathbb N}^{i}+2, \ell}(E_{i})$ 
and~$C_{{\mathbb N}^{i}+2, \ell}(E_{i})$ depend on the oscillator 
quantum~$\hbar\Omega^{i}$
used in the respective NCSM calculations with the maximal number of excitation 
quanta~$N^{i}_{\max}$ as well as the energies $E_i = E_i(N_{\max}^i, \hbar\Omega^i)$.

All resonant states in $^{7}$He can decay via the  $n+{\rm^{6}He}$ channel
with $^{6}$He in the ground state. Additionally, all examined resonances with the exception of the
low-lying $3/2^{-}$ resonance, can decay also via a two-body channel ${n+{\rm^{6}He^*}}$
with $^{6}$He in the excited $2^{+}$ state or via a four-body channel
${n+n+n+{\rm^{4}He}}$.


Within our SS-HORSE--NCSM approach, we start from the NCSM calculations of 
the $^{7}$He  eigenenergies~$E_{i}^{7}$ 
 corresponding to a set 
of pairs of the NCSM basis parameters~$N^{i}_{\max}$ 
and~$\hbar\Omega^{i}$, as 
well as, depending on the channel of interest, 
of the energies~$E_{i}^6$ of the ground state or the lowest $2^{+}$ state of $^6$He or
the ground state energy~$E_{i}^4$ of $^{4}$He 
obtained by NCSM with the same~$\hbar\Omega^{i}$ and the maximal excitation
quanta~$N^{i}_{\max}$ or~$N^{i}_{\max}-1$ depending on the parity 
of the states of interest in $^{7}$He. 
The number of oscillator quanta of the relative motion~${\mathbb N}^{i}$ 
entering Eq.~\eqref{PhaseSSHORSE}
are defined as
\begin{gather}
{\mathbb N}^{i}=N^{i}_{\max}+N^{7}_{\min}-N^{A}_{\min},
\label{bbN}
\end{gather}
where 
$N^{7}_{\min}$ and $N^{A}_{\min}$ are the minimal total oscillator quanta 
consistent with the Pauli principle in $^{7}$He and $^{A}$He, $A=6$ or~4 in the current work. 
The eigenenergies of relative motion~$E_{i}=E^{7}_{i}-E^{A}_{i}$.

In the case of the four-body decay channel,
we use the democratic decay approximation (also known as true four-body scattering or $4\to4$ scattering) 
suggested in Refs.~\cite{JibKr,JibutiEChaYa}. Democratic decay implies a description of the 
continuum using a 
hyperspherical harmonics (HH) basis.
We use here the minimal approximation for the four-body decay mode; i.\,e., only HH with hyperspherical 
momentum~$K= K_{\min}=0$ or~1 for positive or negative parity resonances, respectively, are retained in 
the SS-HORSE extension of the NCSM. This approximation relies on the fact that the decay in the hyperspherical states with $K> K_{\min}$ is strongly suppressed by a large hyperspherical centrifugal 
barrier~$\mathscr{L}(\mathscr{L}+1)/\rho^{2}$, where the effective momentum~$\mathscr{L}=K+3$
and the hyperradius~$\rho^{2}= \sum_{i=1}^{4}(\mathbf{r}_{i}-\mathbf{R})^{2}$, 
$\mathbf{R}$ is the  center-of-mass coordinate, and~$\mathbf{r}_{i}$  are the coordinates of decaying 
neutrons and $^{4}$He. Note that all possible HH are retained in the NCSM basis;  thus the hyperspherical
states with~$K= K_{\min}=0$ are treated as open channels while all the remaining hyperspherical
states with $K> K_{\min}$ are  treated as closed channels. The accuracy of this 
approximation was confirmed in studies of democratic decays in cluster 
models~\cite{Lur11Li,Lur6He,LurAnn,LurSauAr}; it was also utilized in our successful  study of the tetraneutron~\cite{tetran, Shirokov:2018und}.
In this case we should set~$\ell=\mathscr{L}_{\min}=3$ in Eqs.~\eqref{EffRadius}--\eqref{PhaseSSHORSE},
the relation between the HH momentum~$k$ and energy~$E$ can be found in Refs.~\cite{zayMPQT,zaytmf}.

We perform the NCSM calculations with 
various choices of the basis parameters~$N^{i}_{\max}$ 
and~$\hbar\Omega^{i}$ and obtain a set of values of
the effective range function~$K_\ell(E_{i})$ using Eq.~\eqref{PhaseSSHORSE} in some energy interval 
since ${E_i=E_i(N^i_{\max},\hbar\Omega^i)}$. Next we perform a parameterization
of the function~$K_\ell(E)$ which makes it 
possible to calculate the $S$-matrix and its poles 
associated with the resonant states in $^{7}$He.
The effective range function~$K_\ell(E)$ has good analytical 
properties and may be expanded with a Taylor series 
in~$E$~{\cite{Newton}} except for energies in the vicinity 
where the phase shift  takes the values of 0, $\pm\pi$, 
$\pm2\pi$,\:... (this can happen in the resonance region) and the effective range function~$K_\ell(E)$, 
according to Eq.~\eqref{EffRadius}, tends to infinity. Therefore, we
use Pad\'e approximants to parameterize~$K_\ell(E)$;
the number of fit parameters
in the numerator and denominator of the  Pad\'e approximant taken individually in each case
to obtain a reasonable description of selected NCSM
eigenenergies. 


With any set of the Pad\'e approximant parameters
we obtain~$K_\ell(E)$  as a function of energy~$E$ and 
solve  Eq.~\eqref{PhaseSSHORSE} to obtain the 
eigenenergies~$E^{th}_{i}$ which should be obtained in the NCSM 
calculations {with any given combination of~$N^{i}_{\max}$ 
and~$\hbar\Omega^{i}$ to describe exactly this 
function}. These energies~$E^{th}_{i}$ are 
compared with the 
{NCSM eigenenergies~$E_{i}$;}
the optimal values of the fit parameters are found 
by minimizing the sum of squares of deviations 
of~$E^{th}_{i}$ and~$E_{i}$ with weights enhancing the 
contribution of energies obtained with larger~$N_{\max}$ values,
\begin{equation}
\Xi=\sqrt{\frac{1}{p}\sum_{i=1}^p{\left[\left(E_{i}^{th}-
E_{i}\right)^{2}\!\left(\frac{N_{\max}^i}{N_M}\right)^{\!2}\right]}}.
\label{rms_w}
\end{equation} 
Here $ p$ is the number of basis parameter pairs 
and $N_M$ is  the 
largest value of $N_{\max}^i$ used in the fit.


After obtaining 
an accurate parameterization, we express the $S$-matrix through~$K_\ell(E)$ and search 
numerically for the $S$-matrix poles in the complex energy plane as 
described in Ref.~\cite{SSHORSE-K}. 
These poles produce the 
 energies~$E_{r}$ and widths~$\Gamma$ of the $^{7}$He
resonances. 

\section{\boldmath Resonances in $^{7}\mbox{He}$\label{Sec:n-6He}}

We perform the NCSM calculations of $^7$He 
with~$N_{\max}^{i}$ up to~16 for 
negative and up to~17 for positive parity states with 
$\hbar\Omega^{i}$ ranging from 10 to 50~MeV and 
of  $^6$He and $^{4}$He with the same $\hbar\Omega^{i}$ values 
and respective $N_{\max}$ to get the set of relative 
motion eigenenergies~$E_{i}$.  

As stated in Refs.~\mbox{\cite{SSHORSE, SSHORSE_PEPAN, Blkh, 
Blokh, PEPAN2, SSHORSE-K, tetran}}, we cannot use all 
energies~$E_{i}$ 
for the SS-HORSE analysis. In particular, the SS-HORSE 
equations are consistent only with those~$E_{i}$  obtained at any
given~$N_{\max}$ which increase with~$\hbar\Omega$. Therefore, 
from the~$E_{i}$
obtained by NCSM
with any~$N_{\max}$ we should select only those which are 
obtained with~${\hbar\Omega>\hbar\Omega_{\min}}$, 
where~${\hbar\Omega_{\min}}$ corresponds to the 
minimum of the~$E_{i}$.

Next, for the~$K_\ell(E)$  parameterization, we should select only 
the results obtained with large enough~$N_{\max}$ and in the 
ranges of~$\hbar\Omega$ values for each~$N_{\max}$ where the 
continuum state calculations converge, at least, approximately. The 
convergence means that the~$K_\ell(E_{i})$ values [as well as the
respective phase shifts~$\delta_{\ell}(E_{i})$] 
obtained with different pairs of~$N_{\max}^{i}$ and~$\hbar\Omega^{i}$ values form 
a single smooth curve as a function of energy. Our method for the 
selection of the NCSM results is described in detail with a number of
illustrations in Refs.~\mbox{\cite{SSHORSE, SSHORSE_PEPAN, 
Blkh, Blokh, PEPAN2, SSHORSE-K, tetran}}.

\begin{figure}[t]
\centerline{\includegraphics[width=\columnwidth] {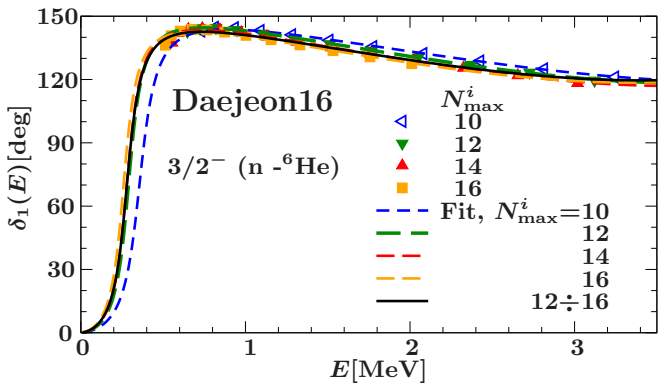}}
\caption{Convergence of phase shifts of the $n{-}{\rm^{6}He}$ scattering in the 
$3/2^{-}$ state in the vicinity of the low-lying resonance in calculations with Daejeon16 $NN$ interaction. 
Symbols are  phase shifts~$\delta_{1}(E_{i})$ at selected energies~$E_{i}$; {curves} are 
\mbox{SS-HORSE} fits to the NCSM results
from different model spaces. Energies are given relative to the $n+{\rm^{6}He}$ threshold.} 
\label{Fig:n-6He_Daej_convergence}
\end{figure}

We illustrate in Fig.~\ref{Fig:n-6He_Daej_convergence}
the convergence in calculations with Daejeon16 $NN$ interaction
of the $n{-}{^6\rm He}$  elastic scattering 
phase shifts for the $3/2^{-}$ state in the vicinity of the low-lying resonance. The energies~$E_{i}$ 
selected from the results of NCSM calculations with~$N^{i}_{\max}=12$, 
14 and~16 generate a set of the phase shifts~$\delta_{1}(E_{i})$ 
shown  by closed symbols which approximately form a single smooth 
curve. The
 SS-HORSE parameterization of  these  3 sets of $3/2^{-}$ phase 
shifts~$\delta_{3}(E_{i})$ (solid curve labeled~$12{\div}16$) accurately describes 
them.
 
We present in Fig.~\ref{Fig:n-6He_Daej_convergence} also the
parameterizations fitted to the NCSM eigenenergies from the 
selection obtained individually with each of these three
$N_{\max}^{i}$ values. 
These three parameterizations nearly 
coincide up to
3.1~MeV which is 
the largest of the NCSM eigenenergies corresponding to the $^{7}$He ground state included in the fit. In 
particular, these parameterizations are nearly 
indistinguishable
in the resonance region. As a result,  we obtain very
similar
resonance energies~$E_{r}$ and widths~$\Gamma$ with these three 
parameterizations\- (see 
Table~\ref{Tab:convergence32mDaej}).

To further
elucidate the convergence trends, we present in 
Fig.~\ref{Fig:n-6He_Daej_convergence} also the phase shifts 
obtained from the NCSM results with~$N^{i}_{\max}=10$ 
and~$\hbar\Omega^{i}$ ranging from 15~MeV to 40~MeV together with the 
respective parameterization. We do not include these
$N^{i}_{\max}=10$ results 
in our selection of the NCSM eigenenergies
used in Eq.~\eqref{rms_w} since 
the respective 
phase shifts~$\delta_{1}(E_{i})$ show more significant deviations from the common curve 
formed by the NCSM results in the three
larger model spaces at the energies 
above the resonance region. 
However, the deviation of these parameterized~$N_{\max}^{i}=10$ phase shifts  from those obtained in larger model 
spaces is not large 
in the resonance region, 
which is of our primary interest.
As a result, the~$N_{\max}^{i}=10$ resonance parameters (see 
Table~\ref{Tab:convergence32mDaej}) 
are within 30\% of those 
obtained in larger model spaces.

We use the spread of the results presented in 
Table~\ref{Tab:convergence32mDaej} (excluding those obtained 
with~$N_{\max}^{i}=10$)~to evaluate the uncertainties of the obtained 
resonance and low-energy scattering parameters. To justify these 
uncertainties, we perform also a few alternative selections 
of {the NCSM} energies~$E_{i}$, e.\:g., we reduce the set of 
selected NCSM energies obtained with~$N_{\max}^{i}=12$ and~14 by 
excluding the eigenstates above the resonant region or extend it by
adding the results of calculations with additional~$\hbar\Omega^{i}$ values 
producing the phase shifts~$\delta_{1}(E_{i})$ which deviate more from the common curve~$12{\div}16$.
Performing the phase shift parameterizations with 
these energy selections for~${N_{\max}^{i}=12}$ and~14 individually
as well as parameterizing\- all these~$N^{i}_{\max}=12$ and~14 results 
together with previously selected~${N^{i}_{\max}=16}$ energies, we 
obtain the spreads of the  $^{7}$He resonance parameters within 
the ranges shown in Table~\ref{Tab:convergence32mDaej}.


\begin{table}[!t]
\vspace{-2.8ex}
\caption{Convergence  with increasing $N^i_{\max}$ of energy~$E_r$ 
(relative to the $n+{^6\rm He}$ threshold) 
and width~$\Gamma$
of the low-lying  resonance $3/2^-_{1}$ in 
$^{7}$He in the $n+{\rm^{6}He}$ channel %
in calculations with Daejeon16 $NN$ interaction.}
\vspace{1.ex}
\begin{ruledtabular}
\begin{tabular}{c|ccccc}
$N^i_{\max}$ & 10    & 12    & 14    & 16  &12$\div$16\\
\hline
$E_r$, MeV   & 0.356 & 0.289 & 0.279 & 0.259 & 0.279 \\
$\Gamma$, MeV& 0.155 & 0.127 & 0.127 & 0.123 & 0.131  \\
\end{tabular}
\label{Tab:convergence32mDaej}
\end{ruledtabular}
\end{table}

\begingroup
\begin{table*}[!t]
\vspace{-2.3ex}
\caption{Energies $E_r$ (relative to the $n+{\rm^{6}He}$ threshold)
 and widths~$\Gamma$ of resonant states in $^{7}$He obtained with JISP16 and Daejeon16
 in the channels $n{-}{\rm^{6}He}$ and $n{-}{\rm^{6}He^{*}(2^{+})}$ and our final predictions 
 based on combining the results in these individual channels.
Estimates of uncertainties of the quoted results 
are presented in parentheses. Results from 
GSM~\cite{GSM}, G-DMRG~\cite{DMRG},  \mbox{CS-COSM}~\cite{CS-COSM} (only widths),  
NCSMC~\cite{Navratil7He1, Navratil7He2}  and NCSMch~\cite{Rodkin-Tchu}
calculations (in the NCSMch case the width in the line ``Predictions'' is obtained by summing  
widths in individual channels) together 
with experimental data are shown for comparison. All values are 
in MeV unless other units are specified.}
\label{Tab:resonance}
\vspace{1.4ex}
\begin{ruledtabular}
%
\begin{tabular}{ccc|cc|ccccc|c}
 \multicolumn{3}{c|}{Resonance}&\multicolumn{2}{c|}{This work} & \multicolumn{5}{c|}{Other theoretical works} & Experiment \\
$J^{\pi}$($^7$He)&$J^{\pi}$($^6$He)& &\parbox{10ex}{JISP16}    & Daejeon16 & \parbox{10ex}{\mbox{NCSMch}}   & \parbox{9ex}{NCSMC}  & GSM     & G-DMRG  & CS-COSM &        \\
\hline
$3/2^-_{1}$      & $0^+$    &  \raisebox{0pt}[3.4ex][2.4ex]{\parbox{2.9ex}{$E_{r}$\\$\Gamma$}}     &  \parbox{10ex}{0.665(12)\\$0.57(4)$}   &  \parbox{10ex}{0.28(4)\\$0.13(2)$}  & \parbox{10ex}{0.547\\ 0.334}   & \parbox{5ex}{0.71\\ 0.30}   & \parbox{5ex}{0.39\\ 0.178}    & \parbox{10ex}{0.460(7)\\ 0.142}& \parbox{10ex}{\mbox{\vphantom{3}}  \\0.048}        & \parbox{10ex}{0.430(3) 0.182(5)}   \cite{Cao} \\
\hline
$1/2^{+}$        & $0^+$      &  \raisebox{0pt}[3.4ex][2.4ex]{\parbox{2.9ex}{$E_{r}$\\$\Gamma$}}    &           &           & \parbox{10ex}{1.696\\ 2.670}  &        &                  &         &            &            \\
\hline
$1/2^-$          & $0^+$     &    \raisebox{0pt}[3.4ex][2.4ex]{\parbox{2.9ex}{$E_{r}$\\$\Gamma$}}      & \parbox{10ex}{2.7(8)\\5.0(6)}    & \parbox{10ex}{2.7(4)\\4.3(3)}    & \parbox{10ex}{2.318\\2.071}  & \parbox{5ex}{2.39\\2.89}   &         & \parbox{10ex}{1.811(6)\\2.150}&  \parbox{10ex}{\mbox{\vphantom{3}}  \\ 2.77}      & \parbox{5.9ex}{3.0(5)\\2}   \cite{Wuosmaa};  \parbox{3ex}{3.5\\10}   \cite{Boutachkov}; \parbox{7.ex}{1.0(1) \\0.75(8)}    \cite{Meister}\\
\hline
$5/2^{-}$        & $0^+$     &   \raisebox{0pt}[3.4ex][2.4ex]{\parbox{2.9ex}{$E_{r}$\\$\Gamma$}}        & \parbox{10ex}{4.4(4)\\1.56(4)}   & \parbox{10ex}{3.63(16)\\1.36(3)}  &  \parbox{10ex}{3.437\\$52$~eV}               &         &         &         &            &            \\
                 & $2^+$  &   \raisebox{0pt}[3.4ex][2.4ex]{\parbox{2.9ex}{$E_{r}$\\$\Gamma$}}   & \parbox{10ex}{3.85(15)\\2.5(2)}  & \parbox{10ex}{3.23(25)\\2.28(8)}  &\parbox{10ex}{3.437\\ 1.941}        &\parbox{5ex}{3.13 \\1.07} &         &         &            &            \\
                  \multicolumn{2}{c}{Predictions} &  \raisebox{0pt}[3.4ex][2.4ex]{\parbox{2.9ex}{$E_{r}$\\$\Gamma$}}   &\parbox{10ex}{4.1(7)\\2.0(7)} & \parbox{10ex}{3.4(4)\\1.8(5)}    & \parbox{10ex}{3.437\\ 1.941}   &\parbox{5ex}{3.13 \\1.07} & \parbox{6ex}{3.47(2)\\2.3(3)} & \parbox{10ex}{3.311(2)\\1.726}& \parbox{10ex}{\mbox{\vphantom{3}}  \\1.80}        & \parbox{10ex}{3.36(9)\\1.99(17)}   \cite{Tilley}\\
\hline
$3/2^-_{2}$      & $0^+$     &  \raisebox{0pt}[3.4ex][2.4ex]{\parbox{2.9ex}{$E_{r}$\\$\Gamma$}}     &  \parbox{10ex}{5.8(5)\\4.11(23)}    & \parbox{10ex}{5.0(3)\\2.84(24)}    &\parbox{10ex}{3.921\\0.229}   &                 &         &         &            &            \\
                 & $2^+$    & \raisebox{0pt}[3.4ex][2.4ex]{\parbox{2.9ex}{$E_{r}$\\$\Gamma$}}   & \parbox{10ex}{5.3(4)\\3.9(6)}    & \parbox{10ex}{4.4(4)\\ 3.9(3)}   &\parbox{10ex}{3.921\\1.459}  &     &     &     &     &            \\
                   \multicolumn{2}{c}{Predictions}  & \raisebox{0pt}[3.4ex][2.4ex]{\parbox{2.9ex}{$E_{r}$\\$\Gamma$}}  & \parbox{10ex}{5.6(7)\\4.0(7)}    & \parbox{10ex}{4.7(7)\\3.4(8)}   & \parbox{10ex}{3.921\\1.796}  &        &         &         &    \parbox{10ex}{\mbox{\vphantom{3}}  \\2.29}      &                        \\
\hline
$3/2^{+}$        & $0^+$     &  \raisebox{0pt}[3.4ex][2.4ex]{\parbox{2.9ex}{$E_{r}$\\$\Gamma$}}  & \parbox{10ex}{$6.5(1.6)$\\$5.9(1.0)$} & \parbox{10ex}{$3.9(6)$\\$4.2(7)$}   &\parbox{10ex}{3.492\\$83.4$~keV}    &                &         &         &            &            \\
                 & $2^+$    & \raisebox{0pt}[3.4ex][2.4ex]{\parbox{2.9ex}{$E_{r}$\\$\Gamma$}}   &   &    &\parbox{10ex}{3.492\\2.508}   &                &         &         &            &            \\
    \multicolumn{2}{c}{Predictions} &  \raisebox{0pt}[3.4ex][2.4ex]{\parbox{2.9ex}{$E_{r}$\\$\Gamma$}}   &\parbox{10ex}{$6.5(1.6)$\\$5.9(1.0)$} & \parbox{10ex}{$3.9(6)$\\$4.2(7)$}   &\parbox{10ex}{3.492\\2.591}&        &         &         &         &                        \\
\hline
$5/2^{+}$        & $0^+$     & \raisebox{0pt}[3.4ex][2.4ex]{\parbox{2.9ex}{$E_{r}$\\$\Gamma$}}     & \parbox{10ex}{6.7(1.5)\\5.8(8)}  & \parbox{10ex}{3.7(7)\\4.4(9)}    & \parbox{10ex}{3.564\\0.258}   &        &         &         &         &                       \\
                & $2^+$    & \raisebox{0pt}[3.4ex][2.4ex]{\parbox{2.9ex}{$E_{r}$\\$\Gamma$}}   &   &    &\parbox{10ex}{3.564\\2.251}  &        &         &         &         &                       \\
   \multicolumn{2}{c}{Predictions} &  \raisebox{0pt}[3.4ex][2.4ex]{\parbox{2.9ex}{$E_{r}$\\$\Gamma$}}   &\parbox{10ex}{6.7(1.5)\\5.8(8)}  & \parbox{10ex}{3.7(7)\\4.4(9)}    & \parbox{10ex}{3.564\\2.512}&        &         &         &         &                       \\
\end{tabular}
\end{ruledtabular}
\end{table*}
\endgroup

\begin{figure*}[!]
\includegraphics[width=\columnwidth]{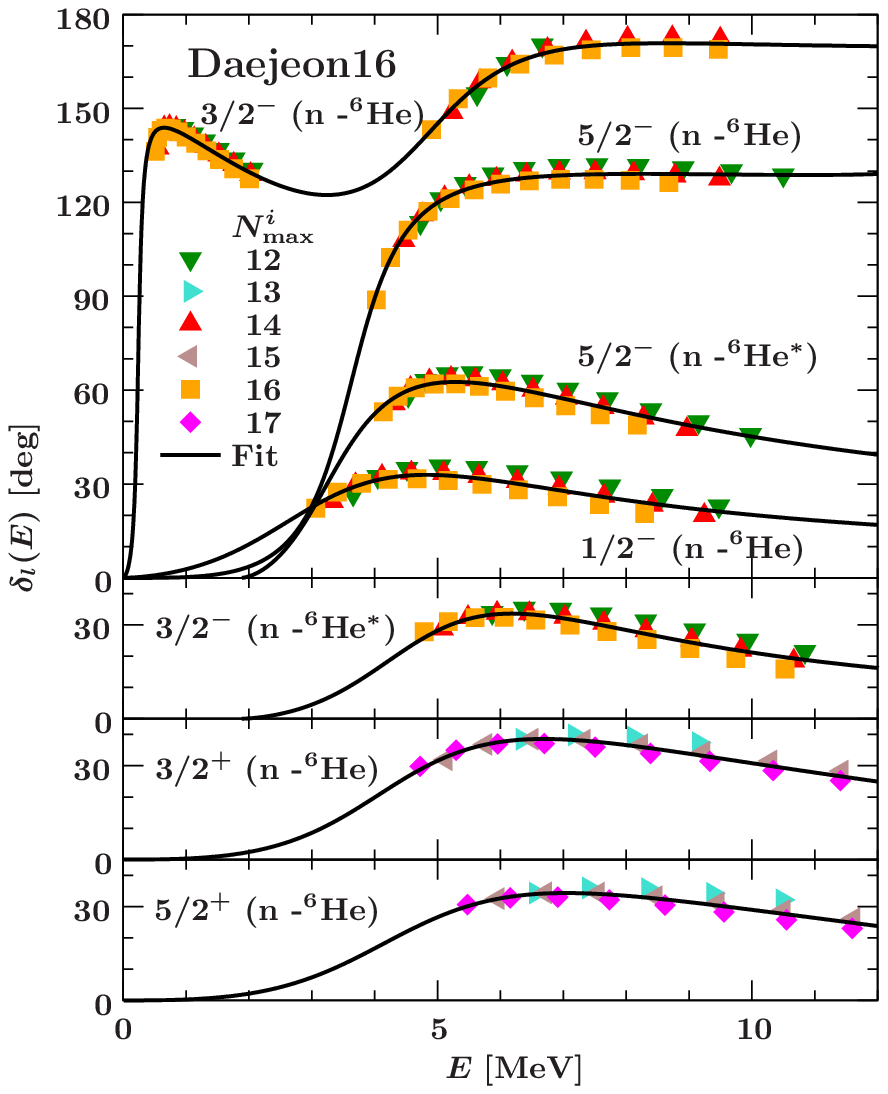}\hfill
\includegraphics[width=\columnwidth]{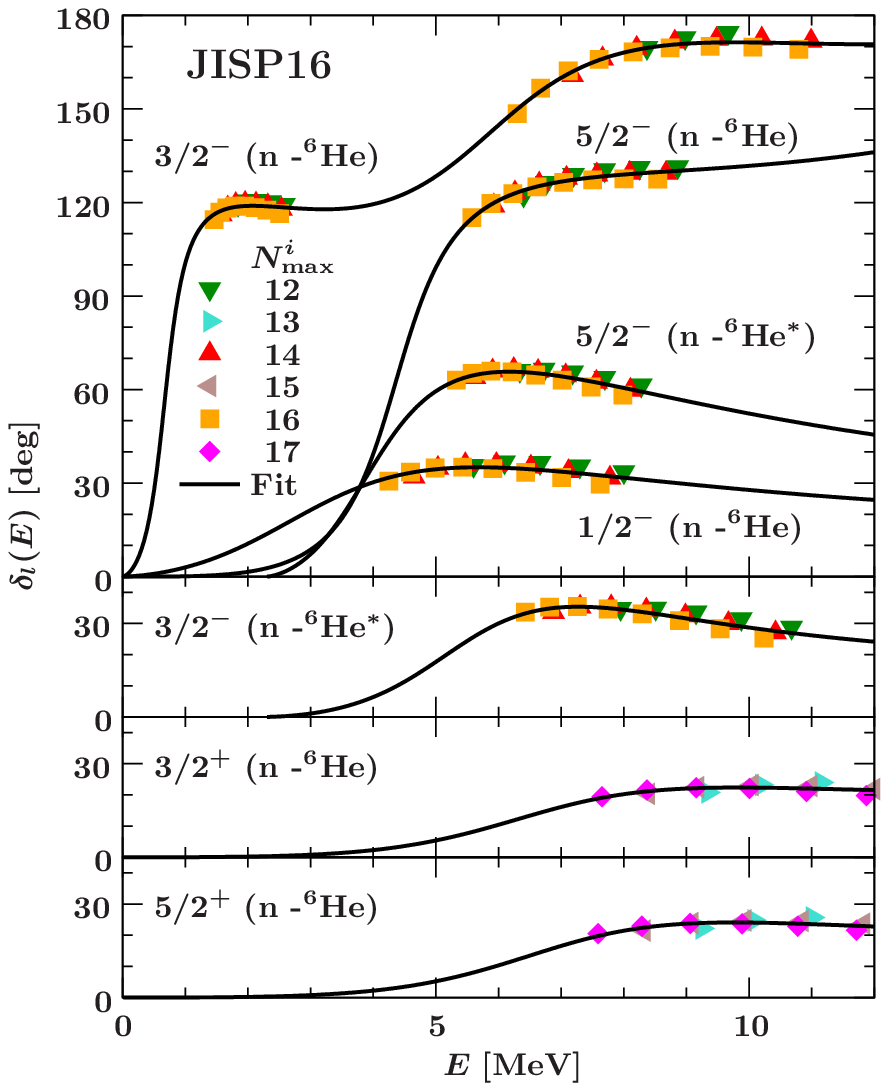}
\caption{Phase shifts in the {$n{-}{\rm^{6}He}$ and in some $n{-}{\rm^{6}He^{*}}$ channels} with the 
Daejeon16 (left) and JISP16 (right) $NN$ interactions. See Fig.~\ref{Fig:n-6He_Daej_convergence} for details.}
\label{Fig:n-6He_Daej}
\end{figure*}

We adopt the same approach for the studies of all resonances in each channel. That is we use the 
results of the NCSM calculations in the three largest model spaces, calculate the resonance energy and
width  for each of these model spaces individually and for the combination of all selected results from
these model spaces and vary the ranges of energy selections in these model spaces to obtain the spreads
of resonance energy and width; these spreads are used as uncertainty estimates while their central values are
used as predictions for the energy and width.   These predictions together with their uncertainties
for various $^{7}$He resonances
based on our calculations with Daejeon16 and JISP16 $NN$ interactions in two-body decay channels
are summarized  in Table~\ref{Tab:resonance}.

 For comparison, we present in 
Table~\ref{Tab:resonance} also available resonance parameters from the
studies within NCSMch~\cite{Rodkin-Tchu},
 NCSMC~\cite{Navratil7He1, Navratil7He2}, GSM~\cite{GSM}, G-DMRG~\cite{DMRG}
and CS-COSM~\cite{CS-COSM} (only widths which are given by numbers in Ref.~\cite{CS-COSM};
the energies in this paper are shown only in figures where they are seen to be close to the
respective experimental values). Note that all these other theoretical calculations where performed 
with different interactions with an exception of the NCSMch studies of Ref.~\cite{Rodkin-Tchu} where
the Daejeon16 was employed.

The low-lying  $3/2^-_{1}$ 
resonance should be clearly related to the experimental 
$3/2^{-}$ resonance in $^{7}$He. Daejeon16
underestimates while JISP16 overestimates
 both its energy and width as compared with experiment;
the NCSMC overestimates these resonance parameters
while  better estimates of 
this resonance were obtained in the GSM and G-DMRG studies. 
We note, however, that, contrary to our NCSM and the NCSMC {\em ab initio} calculations, both of which employ realistic $NN$ interactions,
the GSM and G-DMRG approaches utilize phenomenological $n{-}{\rm^{4}He}$ and
$NN$ interactions fitted to spectra of light nuclei.

It is interesting to compare our results with Daejeon16 with those of the NCSMch studies.
The NCSMch energy and width of this resonance are respectively nearly twice
and three times larger than ours. The NCSMch approach of Ref.~\cite{Rodkin-Tchu} 
is also based on the NCSM calculations (though in smaller model spaces) and utilizes the same 
Daejeon16 $NN$ interaction. However the resonance parameters are obtained within NCSMch in a very different manner. In particular, the NCSMch resonance energy is obtained using a phenomenological
exponential extrapolation A5~\cite{Shin:2016poa}. This extrapolation was designed for the bound states and
has never been applied to resonances before the investigations of Ref.~\cite{Rodkin-Tchu}. There are various
phenomenological exponential extrapolations on the market, all of them are known to provide similar 
results. They are sometimes used to estimate resonance \mbox{energies}, in particular, our group was 
exploiting exponential extrapolations of Ref.~\cite{extra} in the studies of resonances, e.\;g., in
Refs.~\cite{14F,NN-NNN-book}. However, the applicability of the exponential extrapolations to
estimation of resonant energies 
 is unclear. 
We have shown in Ref.~~\cite{inverseNA} that the NCSM eigenstates can differ essentially from the
resonance energy when the resonance width is comparable to its energy.
The resonance
width is obtained within the NCSMch by matching the NCSM wave function with the respective channel
wave function. The resulting width depends strongly on the resonance energy. For example, the authors
of Ref.~\cite{Rodkin-Tchu} mention that switching from their extrapolated energy of the $3/2^-_{1}$ 
resonance of 547~keV to the experimental energy of 430 keV results in the change of the
width $\Gamma=334$~keV to $\Gamma=250$~keV. Thus the accuracy of the NCSMch results for this
and other resonances is unclear, the authors of Ref.~\cite{Rodkin-Tchu} do not report an estimate for it.

We note that our SS-HORSE approach to calculations of resonance energies and widths
was carefully tested and justified in Refs.~\cite{SSHORSE,Blkh,Blokh,SSHORSE-K}
using model problems where resonant parameters were calculated using
other methods. Also, in contrast with the NCSMch, we present the estimates of our uncertainties,  
and we suggest that our predictions obtained with the same interaction to be reliable within
our quoted uncertainties.

The $3/2^{-}$ phase shifts $\delta_{1}(E_{i})$ discussed in Fig.~\ref{Fig:n-6He_Daej_convergence} 
are given in the vicinity of the lowest $3/2^{-}_{1}$ resonance and the symbols in this figure correspond
to the NCSM results for the $^{7}$He ground state. Combining them with those corresponding 
to the $3/2^-$ first excited state,
we obtain the phase shifts revealing  two resonances. The $3/2^{-}$ phase shifts 
in a larger energy scale together with other
phase shifts in various two-body channels obtained with both JISP16 and Daejeon16 interactions are shown 
in  Fig.~\ref{Fig:n-6He_Daej}. (The phase
shift parameterizations shown in Fig.~\ref{Fig:n-6He_Daej_convergence} and resonant parameters
presented in Table~\ref{Tab:convergence32mDaej} were obtained by the fit to both the ground and first
excited states.) The higher lying 
$3/2^{-}_{2}$ resonance can decay not only via the  $n{-}{\rm^{6}He}$ but also via the $n{-}{\rm^{6}He^{*}}$ channel. The $3/2^-$ phase shifts  in this second channel as well as phase shifts in some other states in 
the $n{-}{\rm^{6}He^{*}}$  channels
 are also presented in Fig.~\ref{Fig:n-6He_Daej}. 
The energies and widths of the higher lying 
$3/2^{-}_{2}$ resonance obtained in the $n{-}{\rm^{6}He^{*}}$  channel are close to those obtained
in the $n{-}{\rm^{6}He}$   channel (see Table~\ref{Tab:resonance}).  
Our final prediction for the
energy and width of the $3/2^{-}_{2}$ resonance and their uncertainties 
presented in Table~\ref{Tab:resonance} in the line ``Predictions'' are obtained by combining
their spreads in different model spaces in both channels.

We note that the SS-HORSE extension of the NCSM
opens only a single decay channel of the  resonance; however, all the remaining channels are
still present in the NCSM calculation of $^{7}$He and are coupled to the specified open channel as
closed channels. In a multi-channel calculation, one will obtain the poles at the same location in all matrix
elements of the $S$ matrix. The comparable locations of the $S$ matrix poles obtained with different 
open channels within the  SS-HORSE approach 
confirms the validity of our treatment of the resonance. 
The obtained resonance widths are total widths of  resonances associated with decay in all possible channels.

Within the NCSMch approach, the energy of the $3/2^{-}_{2}$ resonance is obtained by extrapolating
the first excited $3/2^{-}$ state obtained in the NCSM calculations which is independent from the decay 
channel and hence it appears the same in all open channels. However the width obtained by matching
the NCSM wave function with the scattering wave function,   depends strongly
on the considered channel and has a meaning of the partial width characterizing  a probability of 
the decay in the respective channel. Therefore the NCSMch widths are very different in different channels
and should not be compared with ours in each channel. In the line ``Predictions'' we present the NCSMch
result for the total width by summing their widths in individual channels. These total NCSMch widths can
be compared with ours.

Our  results for the $3/2^{-}_{2}$ resonance show that this resonance 
 is one of the candidates for the description of the 
experimentally observed resonance of unknown spin-parity at the 
energy of {6.2}~MeV with the width of 4~MeV. With Daejeon16 we 
obtain slightly smaller than experimental values  for
both its energy and width while JISP16 suggests energy and width closer to the experiment. 
This resonance has been studied theoretically before
within the CS-COSM approach where its width was estimated approximately 30--40\% smaller
than in our calculations. The NCSMch predicts the width of this resonance that is more than 2 times smaller
than ours; the NCSMch energy of this resonance is more than 1~MeV smaller than ours.

The $1/2^+$ scattering  phase shifts are found to decrease monotonically with energy
 without any signal of a resonant state in calculations with both Daejeon16 and JISP16 interactions. This result is 
in an agreement with the experimental data and the GSM predictions 
of Ref.~\cite{GSM} and the NCSMC predictions of 
Refs.~\cite{Navratil7He1, Navratil7He2}. From our parameterization 
of the effective range function~$K_\ell(E)$ we obtain the 
scattering length $a_0=2.2(4)$~fm and the effective radius
$r_0=2.1(1.1)$~fm for the $n+{\rm^{6}He}$ $s$-wave scattering.

The NCSMch studies of Ref.~\cite{Rodkin-Tchu} propose the $1/2^+$ resonance at the
energy~$E_{r}=1.696$~MeV with the width~$\Gamma=2.670$~MeV. We note here that, as was
clearly demonstrated in Ref.~\cite{inverseNA}, not all NCSM eigenstates should be associated with a
resonance.  Furthermore, the non-resonant scattering requires an appearance of some NCSM eigenstates for
compatibility with the respective phase shifts. However any NCSM eigenstate with positive energy with
respect to any threshold can be matched with any open channel thus producing a theoretical prediction for a resonance which may not correspond to a physical resonance. We suppose
this is a drawback of the NCSMch approach which is in particular manifested in the case of the spurious
$1/2^+$ resonance. We should note however that, according to the NCSMch predictions, the width of the 
$1/2^+$ resonance  is much larger than its energy; thus this resonance will not be
pronounced in a scattering experiment though may be detected in other 
reactions.  This latter situation seems to occur in the case of the tetranuetron resonance where theory~\cite{tetran} and some experiment~\cite{tetran-exp} suggest its width is larger than its energy.

 The results for the $1/2^-$ resonance presented in
Table~\ref{Tab:resonance}, contrary to the  $3/2^{-}_{2}$ resonance, were obtained only in the
channel   $n{-}{\rm^{6}He}$ with the  ${\rm^{6}He}$ in the ground state. This resonance 
with the width of approximately 4~MeV or more should have the energy less than 1~MeV in 
the $n{-}{\rm^{6}He^{*}}$ channel. Clearly, the $n{-}{\rm^{6}He^{*}}$ phase shift will be nearly unaffected by
the respective $S$ matrix resonant pole and hence it is not feasible to deduce the pole location from
these phase shifts. We obtain the same energy of this resonance with Daejeon16 and JISP16 interactions
which are slightly larger the results of other theoretical studies. The widths predicted by Daejeon16 and JISP16
are close to each other and approximately twice as large as those reported in other theoretical papers.

The experimental situation for the $1/2^-$ resonance is not clear. 
While the resonant energies of Refs.~\cite{Wuosmaa, Boutachkov} 
are comparable, the widths are very different. The results of our work and other theoretical works for the 
resonance energy are in fair agreement with  the neutron pickup
and proton-removal reaction experiments~\cite{Wuosmaa}. However for the  width of this resonance we 
obtain a value that is approximately two times larger than in
experiment~\cite{Wuosmaa} and approximately two times smaller  
than in experiment~\cite{Boutachkov}. It is clear that our results do not 
support the interpretation of experimental data on one-neutron 
knockout from $^{8}$He of Ref.~\cite{Meister} advocating a 
low-lying ($E_r\sim1$~MeV) narrow ($\Gamma<1$~MeV) $1/2^-$ 
resonance in $^{7}$He.

We obtain very similar results for the $5/2^-$ resonance in the $n{-}{\rm^{6}He}$ and $n{-}{\rm^{6}He^{*}}$  channels 
as well as in calculations with Daejeon16 and JISP16 interactions.
 It may look surprising that we got a wide resonance in the $n{-}{\rm^{6}He}$ 
channel where the orbital momentum~$\ell=3$ produces a high centrifugal barrier. We note again that the
respective $S$ matrix resonant pole appears due to the coupling to other channels and 
provides the information about the total resonance width 
associated with all possible channels and may be very different from the partial width 
associated with the decay probability in one particular channel: the partial width in this channel of 52~eV 
obtained by the NCSMch is 4 orders of magnitude smaller. Our results for the energy and total
width of the $5/2^-$ resonance  are seen to be in good agreement with 
experiment and with the other available theoretical studies performed 
with different interactions and using different approaches.


Our results for the positive parity $3/2^{+}$ and  $5/2^{+}$ resonances presented in Table~\ref{Tab:resonance} 
were obtained only in the  $n{-}{\rm^{6}He}$ channel. Note, these resonances are wide: their  widths  obtained
with Daejeon16 are more than 4~MeV and  are larger than
their energies; their widths obtained with JISP16 are close to 6 MeV and their energies are only slightly larger.
In the $n{-}{\rm^{6}He^{*}}$  channel their energies become  smaller than their 
widths. Therefore the resonances are not well-resolved in this channel and we do not attempt to
extract   resonance 
parameters from the  $n{-}{\rm^{6}He^{*}}$ 
phase shifts.

The  $3/2^{+}$ and  $5/2^{+}$ phase shifts are seen in Fig.~\ref{Fig:n-6He_Daej} 
to nearly coincide and behave very
similar to the $3/2^{-}$ phase shifts in the  $n{-}{\rm^{6}He^{*}}$ channel that is most noticeable in the
case of JISP16. Therefore we 
obtain the $3/2^{+}$ and  $5/2^{+}$ resonances at  energies close to that
of the $3/2^{-}_{2}$ resonance but their widths are slightly larger.
As a result, we suppose that the wide resonance observed
around 6~MeV is formed as a complicated overlap  of
the $3/2^{-}_{2}$, $3/2^{+}$ and  $5/2^{+}$ resonances.
{Note, this wide experimental resonance  overlaps partially also with the $1/2^{-}$ and $5/2^{-}$ resonances.}

The  $3/2^{+}$ and  $5/2^{+}$ resonances in the $n{-}{\rm^{6}He}$ channel are characterized by the
orbital momentum $l=2$ or higher. Our phase shifts reflect the pole structure of the $S$ matrix. However
the orbital momentum $l=2$ suggests a high centrifugal barrier. Therefore the partial widths
of these resonances in the $n{-}{\rm^{6}He}$ channel are suppressed as is manifested in the NCSMch
calculations. The total width of the  $3/2^{+}$ and  $5/2^{+}$ resonances in the NCSMch is dominated by 
contributions from other channels and appears to be much smaller. The energies of these resonances
deduced in the NCSMch by exponential extrapolations is also smaller than our predictions with the same 
interaction and this distinction seems to be a common feature for all wide resonances reported here. As a result, 
as seen in Table~\ref{Tab:resonance}, the NCSMch predicts
comparable  energies and similar total widths for  the $5/2^{-}$, $3/2^{-}_{2}$, $3/2^{+}$, and $5/2^{+}$
resonances in $^{7}$He. In other words, according to the NCSMch, the resonance in $^{7}$He at the energy
of 3.36~MeV with the width of approximately 2~MeV which spin-parity has a  tentative assignment  
of $5/2^{-}$, appears as a complicated overlap of $5/2^{-}$, $3/2^{-}_{2}$, $3/2^{+}$, and $5/2^{+}$
resonances while, at the same time, there is no NCSMch indication of the wide resonance around 6~MeV.

We also examined the democratic  four-body ${\rm^{4}He}+n+n+n$ decay channels of all $^{7}$He resonances with the 
exception of the lowest
$3/2^{-}$ resonance which is below the respective threshold. In all cases we obtain  resonances with
energies close to those of respective two-body channels 
but with much smaller widths --- at least 3 times smaller and sometimes more than an order of magnitude smaller. We conclude that the  direct democratic  four-body decays of $^{7}$He resonances are
suppressed due to the large hyperspherical centrifugal barrier~$\mathscr{L}(\mathscr{L}+1)/\rho^{2}$ 
which dynamically pushes the system to form $^{6}$He in the ground or excited 
resonant $2^{+}$ states in the decay process. 
Therefore we do not present the results for these direct democratic decays of 
$^{7}$He resonances in Table~\ref{Tab:resonance} and in the figures. 
{The democratic decay channels should be treated separately and
cannot be included in a multi-channel calculation together with two-body
decay channels since the democratic and two-body decay channel wave functions cannot be orthogonalized.}
We note, however,
that the four-body ${\rm^{4}He}+n+n+n$ decays of $^{7}$He resonances occur as two-step
processes in the $n{-}{\rm^{6}He^{*}}$ channels when $^{7}$He first emits  a neutron leaving the
excited ${\rm^{6}He^{*}}$ $2^{+}$ state which emits  $^{4}$He and two neutrons.


\section{Summary and conclusions}

Motivated by experimental uncertainties in the 
properties of the unbound nucleus $^7$He, we solved for 
$^{7}$He resonances using 
the 
SS-HORSE extension of the {\it ab 
initio} NCSM with the realistic 
Daejeon16 and JISP16 $NN$ interactions.

The  four-body  ${{\rm^{4}He}+n+n+n}$ direct democratic decays of $^{7}$He resonances
were found to be suppressed. All examined resonances {may} decay via the $n{-}{\rm^{6}He}$ 
channel with $^{6}$He in the ground state  and, with an exception of the $3/2^-_{1}$ resonance,
via the $n{-}{\rm^{6}He^{*}}$ channel
with ${\rm^{6}He}$ in the excited $2^{+}$ state. That excited 2+ state subsequently decays emitting $^{4}$He
and two neutrons thus resulting in the  four-body  ${{\rm^{4}He}+n+n+n}$   final state.
The resonance energies and widths quoted in Table~\ref{Tab:resonance}
simulate the results of a multichannel calculation;
in particular, the widths are a reasonable approximation to the total widths of the resonances as
confirmed by similar results obtained
with different open channels. 

Our 
predicitions for the low-lying narrow~$3/2^{-}_{1}$ resonance are in reasonable agreement with 
experiment and with results quoted in the GSM~\cite{GSM},  G-DMRG~\cite{DMRG},
NCSMch~\cite{Rodkin-Tchu}, and 
NCSMC~\cite{Navratil7He1, Navratil7He2} theoretical studies. 

The~$1/2^{-}$ resonance is predicted  at the energy in reasonable agreement with the 
NCSMC~\cite{Navratil7He1, Navratil7He2} and NCSMch~\cite{Rodkin-Tchu} calculations  and results 
of experiments of  Refs.~\cite{Wuosmaa, Boutachkov} and 
about 1~MeV higher than suggested by the \mbox{G-DMRG} studies~\cite{DMRG}. 
The width of this resonance is found to be more than 4~MeV which is larger than the width predicted
in the NCSMch, NCSMC, \mbox{G-DMRG}, and {CS-COSM}~\cite{CS-COSM} calculations 
and larger than the experimental width  of Ref.~\cite{Wuosmaa}. 
However, our~$1/2^{-}$ resonance width is less than half the experimental width
of Ref.~\cite{Boutachkov}. Our results as well as those of the above mentioned NCSMch, 
NCSMC, GSM, G-DMRG and \mbox{CS-COSM} calculations disagree
with the indication of a low-lying narrow resonant $1/2^{-}$ state 
suggested in Ref.~\cite{Meister}.

Our predictions for the relatively wide~$5/2^{-}$ resonance are in reasonable agreement with 
experiment and with results quoted in the NCSMC, NCSMch,  GSM, 
\mbox{G-DMRG} and \mbox{CS-COSM} studies. 

We found  a wide $3/2^{-}_{2}$ resonance around the energy of 5~MeV which was
also predicted in the \mbox{CS-COSM} calculations~\cite{CS-COSM} 
as well as wide $3/2^{+}$ and  $5/2^{+}$ resonances at nearby
energies.
Based on our results, it appears reasonable to propose that the observed resonance at the 
energy of~6.2~MeV with the width of~4~MeV of unknown spin-parity
mentioned in the compilation of Ref.~\cite{Tilley} is formed as 
an overlap of
the $3/2^{-}_{2}$ resonance with 
$3/2^{+}$ and $5/2^{+}$ resonances.

We do not find a resonance in the~$1/2^{+}$ state which is consistent with the findings of the  GSM~\cite{GSM}, 
NCSMC~\mbox{\cite{Navratil7He1, Navratil7He2}} studies and with the experimental 
situation. 

%

\subsection*{Acknowledgements}

{We are thankful to Yu.~M.~Tchu\-vil'sky for discussions.}
This work is supported in part  by the 
National Research Foundation of Korea (NRF) grant funded by the 
Korea government (MSIT) (No.~2018R1A5A1025563), 
by the Russian Foundation for Basic Research under Grant
No.~20-02-00357,
by the Ministry of Science and Higher Education of the Russian 
Federation under Project No.~0818-2020-0005,
by the U.S.\ Department of Energy under Grants
No.~DESC00018223 (SciDAC/NUCLEI) and  
No.~DE-FG02-87ER40371, 
by the Rare Isotope Science Project of the Institute for Basic 
Science funded by Ministry of Science and ICT and National 
Research Foundation of Korea (2013M7A1A1075764). 
Computational resources were provided by the National Energy 
Research Scientific Computing Center (NERSC), which is supported 
by the Office of Science of the U.S.\ Department of Energy under 
Contract No.~DE-AC02-05CH11231, and by the National Supercomputing 
Center of Korea with supercomputing resources including technical 
support (KSC-2018-COL-0002). We acknowledge also the Shared Service Center ``Data Center 
of the Far-Eastern Branch of the Russian Academy of Sciences'' for using their resources.

\end{document}